\documentclass{ieeeaccess}
\usepackage{hyperref}
\usepackage{amsmath,amssymb,amsfonts}
\usepackage{algorithmic}
\usepackage{graphicx}
\usepackage{subcaption}
\usepackage{caption}
\usepackage{multirow}
\usepackage{textcomp}
\usepackage{algorithm}
\usepackage{enumitem}

\def\BibTeX{{\rm B\kern-.05em{\sc i\kern-.025em b}\kern-.08em
    T\kern-.1667em\lower.7ex\hbox{E}\kern-.125emX}}

\begin{document}
\history{Date of publication xxxx 00, 0000, date of current version xxxx 00, 0000.}
\doi{10.1109/ACCESS.2017.DOI}

\title{Security in Next Generation Mobile Payment Systems: A Comprehensive Survey}

\author{
\uppercase{Waqas Ahmed}\authorrefmark{1},
    \uppercase{Aamir Rasool}\authorrefmark{2},
    \uppercase{Abdul Rehman Javed}\authorrefmark{1*},
    \uppercase{Neeraj Kumar}\authorrefmark{3},
        \uppercase{Thippa Reddy Gadekallu}\authorrefmark{4},
        \uppercase{ Zunera Jalil}\authorrefmark{1},
        \uppercase{Natalia Kryvinska}\authorrefmark{5*}
}            
 \address[1]{Department of Cyber Security, Air University, Islamabad, Pakistan}
 \address[2]{Institute of Avionics and Aeronautics, PAF Complex, E-9, Air University, Islamabad, Pakistan}
 \address[3]{Department of Computer Science and Engineering, Thapar Institute of Engineering \& Technology, Patiala (Pb.),India}
\address[4]{School of Information Technology and Engineering, Vellore Institute of Technology, Tamil Nadu, India}
\address[5]{Faculty of Management, Comenius University in Bratislava Odbojárov 10, 82005 Bratislava 25, Slovakia}

\markboth
{Author \headeretal: Preparation of Papers for IEEE TRANSACTIONS and JOURNALS}
{Author \headeretal: Preparation of Papers for IEEE TRANSACTIONS and JOURNALS}

\corresp{Corresponding author: abdulrehman.cs@au.edu.pk, natalia.kryvinska@uniba.sk}

\tfootnote{}

\begin{abstract}
Cash payment is still king in several markets, accounting for more than 90\% of the payments in almost all the developing countries. The usage of mobile phones is pretty ordinary in this present era. Mobile phones have become an inseparable friend for many users, serving much more than just communication tools. Every subsequent person is heavily relying on them due to multifaceted usage and affordability. Every person wants to manage his/her daily transactions and related issues by using his/her mobile phone. With the rise and advancements of mobile-specific security, threats are evolving as well. In this paper, we provide a survey of various security models for mobile phones. We explore multiple proposed models of the mobile payment system (MPS), their technologies and comparisons, payment methods, different security mechanisms involved in MPS, and provide analysis of the encryption technologies, authentication methods, and firewall in MPS. We also present current challenges and future directions of mobile phone security.
\end{abstract}

\titlepgskip=-15pt

\maketitle

\section{Introduction}
Cash payment is still monarch in several markets, accounting for more than 90\% of the payments in all almost all the developing countries \cite{verkijika2020affective}. Nowadays, the use of mobile devices by people has increased tremendously. A considerable number of people use mobile phones to perform day-to-day tasks \cite{javed2020alphalogger}. These devices can be used for many tasks, such as making phone calls, web surfing, emailing, gaming, and many other tasks. 

The current research in the area is focused on the usage of mobile phone to perform payment securely. However, mobile systems face different limitations \cite{cimato2001design,kungpisdan2003practical,marvel2001authentication} such as low storage and computation power, due to which they cannot perform heavy encryption operations. Different attacks are reported on mobile devices due to lack of security patches such as spoofing, phishing, malware, and sniffing attacks \cite{wang2016mobile,deep2020survey,iwendi2020keysplitwatermark,rehmanensemble}. In order to effectively design the MPS, these attack scenarios must be considered for safety and security.

\begin{figure*}[h!]
\centering
\includegraphics[width=\textwidth]{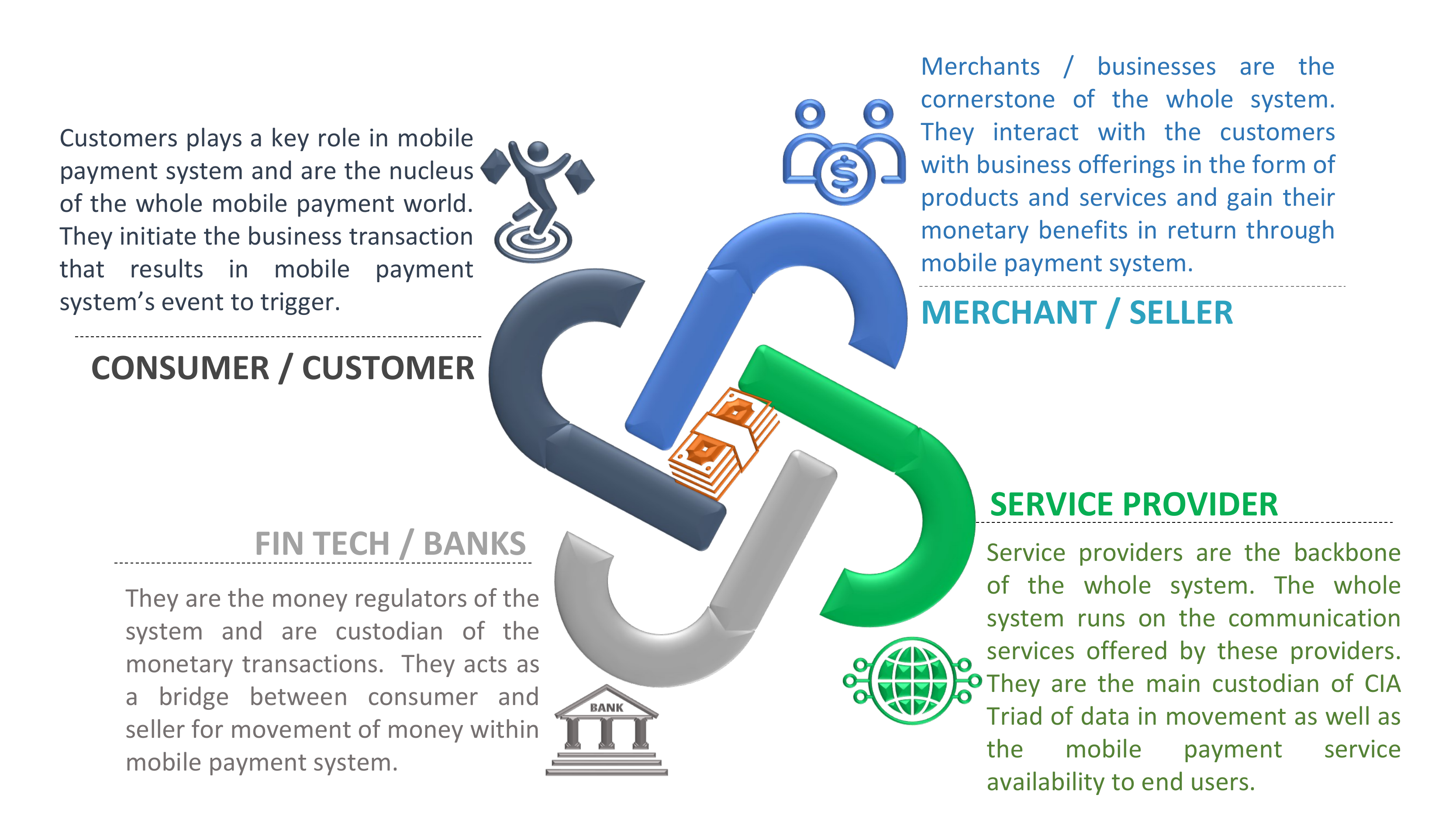}
\caption{Major Components Of MPS}
\label{fig:PTxxx}
\end{figure*}

Information and communication technology (ICT) is being extensively used all around the world \cite{baza2019b}. The traditional face-to-face interaction requirement for payment transactions is avoided, and remote communication is adopted. There is no need for direct contact between a payer and the payee that changes the business environment and leads toward using the internet to do different transactions. This situation requires electronic money or digital bits; the system resembles like traditional payment but with the usage of internet infrastructure and digital data for money transfer. There are many advantages of using e-money, like the client's anonymity or the client's presence is not required during transactions. At the same time, it also has some disadvantages, like compromising of confidentiality, integrity, and availability (CIA) \cite{mohammad2017intelligent}.

The vast development of mobile phone technology enables the growth of internet services. Internet brings the electronic transaction systems \cite{preuveneers2016feature} to the mobile phones and also m-commerce \cite{turban2018mobile} becomes an alternative for e-commerce. As m-commerce is growing at a tremendous pace, it is getting much more attention than e-commerce nowadays. M-commerce has the same characteristics \cite{hubert2017acceptance} as e-commerce with some extra advantages like Mobile Payment System (MPS), that allows clients to perform transactions in real-time by using mobile phones anywhere; all it needs is internet connectivity. Another advantage is that, unlike a PC, one can carry his/her mobile phone anywhere. Some other benefits are interoperability, speed, cost, and cross-border payments. 

\begin{figure*}[h!]
\includegraphics[width=\textwidth]{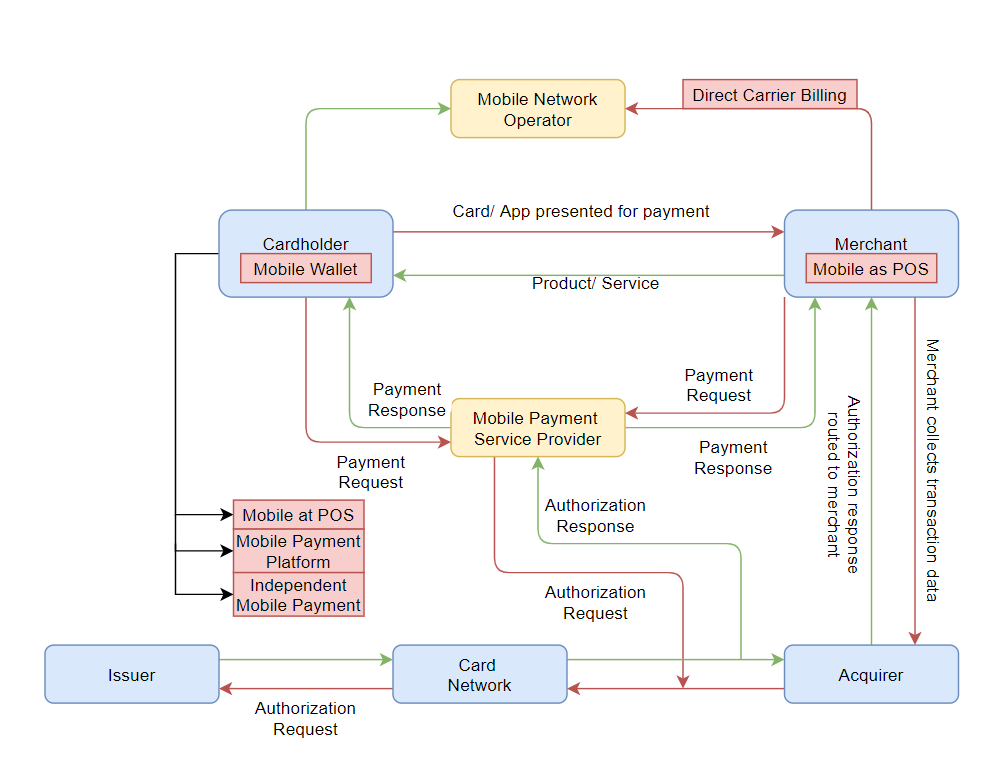}
\caption{Mobile Management System}
\label{fig:MPS}
\end{figure*}

Figure \ref{fig:MPS} shows a Mobile Management system.

A MPS should include authentication, access control, confidentiality, integrity, non-repudiation, and availability \cite{editioncryptography}. 
Authentication process included two steps: verification of the user and verification of the origin. 
In authentication, two processes include verifying the user and the origin of the source of data. Access control can grant access to an authorized person to the payment system and block unauthorized personnel from accessing the payment system. The information must also remain hidden to avoid passive attacks against transaction data. Availability ensures that the payment system is accessible. Integrity avoids the modification of data and non-repudiation ensures that a specific user has transmitted the message. 

Security is essential for MPS, and many security standards such as PCI DSS (Payment Card Industry Data Security Standard) \cite{PCISSC}, which was first released in 2004, is used to maintain the CIA triad. The people or merchants who use payment cards follow PCI DSS standards but security violations can still occur \cite{wang2016mobile,depot2014home}. When security violations occur, personal information, payment card information such as expiration date, ATM card number, security code, and transaction ID are at risk, and it can lead to fraud or illegal usage of service. There are two methods of Mobile Payment Systems: account based payment system and token based payment system \cite{tellez2017mobile}.
\subsubsection{Account Based Payment System}
In the account-based transaction, we need cards or information cards like ATM or credit card. Using this process, the amount is charged from the user's bank account after getting the required details or getting confirmation of the transaction from the user.

\textbf{Risk Factor:} If any misuse of card or details is done or any forgery or identity theft is done, then it will affect this system.

\subsubsection{Token Based Payment System}
It is a new electronic payment method based on tokens instead of cash or credit cards. These tokens are generated by any bank, service provider, or telecom company. Moreover, it is used in the same way as cash is used. By using such tokens, users can pay to any company through mobile, and those tokens will be sent to that company which they can encash, or the provider will pay them for each token.

\textbf{Risk Factor:} These tokens will have no worth if the user has tokens in their account and the merchant does not accept those tokens.

\subsection{Motivation}
\begin{figure}[h!]
\centering
\includegraphics[width=\columnwidth]{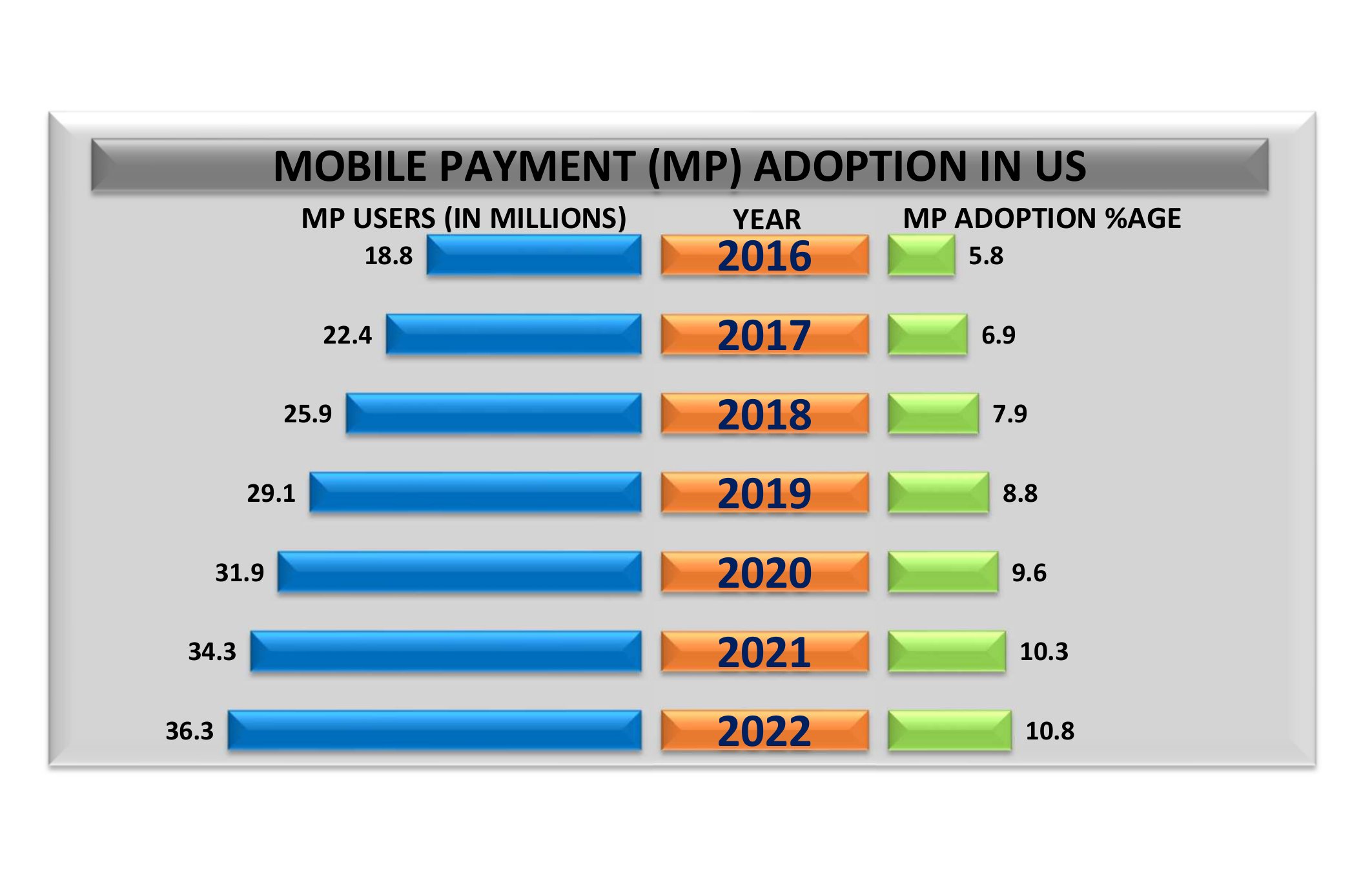}
\caption{Proximity mobile payment adoption in the United States from 2016 onwards}
\label{fig:MPSincrease}
\end{figure}

Mobile phones' usage is highly elevated in the current era compared to their usage a decade before. The number of mobile phones is higher than the number of bank accounts that exist. Due to its high usage level, most business organizations, the entertainment industry, banks, the education sector, and almost all fields turn towards mobile phone adaptability. To benefit from this device, they launch their applications for the comfort of people. Almost all banks facilitate consumers with mobile phone applications. People use mobile phones for shopping, transferring money, and getting various services. The maximum use of mobile devices and versatility motivates us to focus on mobile payment systems (MPSs). Different models of payment systems have been proposed, but many limitations exist; security and privacy concerns. Figure \ref{fig:MPSincrease} shows the increased usage of MPS in the United States from 2016 to onwards and sheds light on the adoption rate and numbers of users (in millions) in a single glance. 
\subsection{Contribution} This research paper aims to present an in-depth analysis and survey of MPS. This paper makes the following contributions:

\begin{enumerate}
\item We present an overview and discussed different components of MPS. We present a review of the existing MPS structure and its limitations.

\item We discuss and analyze the two main methods of MPS included: account-based payment system and token-based payment system.

\item We provide detailed history, development, and deployment of MPS and discussed aspects of MPS included socioeconomic conditions, cost efficiency, diffusion of mobile phones, convenience, new initiatives, heavy restrictions and regulations, limited collaboration, underdeveloped ecosystem, and security problems.

\item We discuss key attributes of MPS, and stakeholder and communication entities' roles in MPS form different aspects.

\item We demonstrate the security mechanisms involved in MPS. Provide analysis of the encryption technologies, authentication methods, and firewall in MPS.

\item We present authentication techniques (one way, two way, and multiple way authentication) in MPS. Discussed the pros and cons of the mentioned techniques. 

\item Provide analysis of the various possible attacks on MPS included attacks against user privacy, attacks on authentication techniques, attacks on user confidentiality data, attacks on the data integrity, and attacks against MPS services availability.

\item Next, we provide key challenges developers and researchers face in implementing and deploying MPS. At the end of the research paper, we present different research directions related to MPS.
\end{enumerate}

\noindent The rest of the paper is organized as follows: section \ref{MPS-History} discusses the history, development, and deployment of the mobile payment system; section \ref{Sec2} discusses the generic architecture of the M-Payment system; section \ref{Sec3} presents technologies used in the M-Payment system and their comparison; section \ref{Sec4} provides security aspects comparison of different mobile payment models; and at the end section \ref{Sec5} provides the current challenges, future direction followed by Section \ref{Sec6} which concludes the works.

\begin{table}[!h]
\centering
\caption{List of Abbreviations}
\label{notation}
\scalebox{0.91}{
\begin{tabular}{|l|l|}
\hline
\textbf{Abbreviation} & \textbf{Description} \\ \hline
MPS & Mobile Payment System\\ \hline
ICT & Information and Communication Technology\\ \hline
CIA & Confidentiality, Integrity and Availability \\ \hline
MMS & Mobile Management System\\ \hline
PCI & Payment Card Industry\\ \hline
DSS & Data Security Standard\\ \hline
PTC & Pakistan Telecommunication\\ \hline
TAM & Technology Acceptance Model\\ \hline
EPS & Electronic Payment System\\ \hline
TE & Technology Evolution\\ \hline
CPT & Cart Present Transactions\\ \hline
SEM & Structural Equation Model\\ \hline
NFC & Near Field Communication \\ \hline
QR & Quick Response\\ \hline
SMS & Short Message Service\\ \hline
MB & Mobile Banking\\ \hline
MP & Mobile Payments\\ \hline
MW & Mobile Wallets\\ \hline
MC & Mobile Currency\\ \hline
MFS & Mobile Financial System\\ \hline
MNO & Mobile Network Operators\\ \hline
DS & Digital Signature\\ \hline
CA & Certificate Authority\\ \hline
RA & Register Authorities\\ \hline
SKE & Symmetric Key Encryption\\ \hline
PKE & Public-Key Encryption\\ \hline
PIN & Personal Identification Number\\ \hline
MFA & Multifactor Authentication\\ \hline
SFA & Single-Factor Authentication \\ \hline
2FA & Two-Factor Authentication\\ \hline
DDOS & Distributed Denial-of-Service Attack\\ \hline
DOS & Denial-of-Service Attack\\ \hline
OTP & One Time Password\\ \hline
USSD & Unstructured Supplementary Service Data\\ \hline
GSM & Global System for Mobile\\ \hline
RFID & Radio Frequency Identification\\ \hline
QRC & Quick Response Code\\ \hline
WAP & Wireless Application Protocol \\ \hline
U2F & Universal 2nd Factor\\ \hline
MAC & Message Authentication Code\\ \hline
SAP & Secure Authentication Protocol \\ \hline

\end{tabular}}
\end{table}

\section{Related Work}
Cash payment is still monarch in several markets, accounting for more than 90\% of the payments in all almost all the developing countries \cite{ verkijika2020affective}. Therefore, it is essential to realize the importance of MP acceptance. Different researchers have completed several research studies on MP after the first PM transaction performed in 1997 \cite{ dahlberg2015critical, dahlberg2008past}. Several studies on MP implementation have the focus to work on the user side. Considering the user's behavior on MP is significant to advance MP services to improve users acceptance intention \cite{lee2019study}.

 \cite{saxena2019survey} tried to respond to certain questions related to the security of online payment systems and presents several ways to overwhelmed different security threats associated with online payment systems. \cite{thangamuthu2020survey} presents different types of online payments such as credit card, e-wallet, debit card, net banking, smart card, mobile payment, and amazon pay. The authors also present some requirements for online payments such as integrity and authorization, out-band authorization, password authorization, signature authorization, confidentiality, and availability and reliability. \cite{saranya2021efficient} proposed a new secure authentication protocol (SAP) for mobile payment. The author used cryptography techniques for the authentication between server and client. The proposed technique provides security to user data account and provides privacy during the payment transaction. 

This research work reviews the literature work of MP in the following significant areas: mobile payment system (MPS): history, development, and deployment; factors limiting MP development; MPS key attributes; and MP stakeholders and entities. Table \ref{ResearchComparision} presents the comparison of the existing survey papers on MP. \\
\newpage
\begin{table*}[!h]
\centering
\caption{Research Literature Comparison }
\label{ResearchComparision}
\scalebox{0.76}{
\begin{tabular}{|p{1.5cm}|p{8cm}| p{12cm}|}
\hline
\textbf{Ref.}& \textbf{Proposed} & \textbf{Outcomes} 
\\ \hline

 \cite{wang2020exploring} 
& To integrate Choice-Based Conjoint (CBC) and System Dynamics (SD), the author develops an empirical data-driven simulation methodology to analyze several competing diffusion dynamics Mobile Payment platform.
& 
To collect multi-attribute preference data, a choice-based conjoint analysis methodology is used on a different platform. To evaluate the effect of platform design strategies, an empirical user preference SD simulation model is developed with the help of an empirical user preference data model.
\\ \hline
 \cite{liebana2017intention} 
& The author in this study aims to compare different factors that define consumers' acceptance for near field communication (NFC) and short message service (SMS) as examples of means for mobile payment of future payment systems.
& From the most relevant literature review, a model is driven and used in this research that applies the perceived security (PS) and technology acceptance model (TAM). The results succeeded in this research work in determining the differences between the different factors that define the acceptance of the Mobile Payment Systems and the intention to use level of the users. 
\\ \hline 
 \cite{saxena2019survey} 
& The author discussed the data security which the application user has shared during online payments. 
& In this research paper, the author answers specific questions related to online payments security and discussed different ways to reduce the security threats associated with online payments. 
\\ \hline 
 \cite{masihuddin2017survey} 
& The paper investigates and increases awareness associated with different electronic payment systems (EPS), including security considerations, challenges, and advantages.
& The author conducted a detailed survey regarding all aspects of EPS after analyzing the existing research studies related to online payment systems. 
\\ \hline 
 \cite{lee2019study} 
& The author discussed distributed systems and their regulatory compliance related to this approach which is not decentralized yet.
& For instant payment systems, the author reviewed some distributed protocols to find the possibility to syndicate distributed systems with centralized services effectively in the mobile payment system. 
\\ \hline
 \cite{kumar2020structural} 
& With the case of Mobile Payment Technology, the author focus on identifying different technological trajectories in the ecosystem (technological).
& The novelty in this research is finding the technological evolution with the help of patent citation and social network analysis. The case of the Mobile Payment Technology ecosystem is analyzed quantitatively. This research aims to provide a path to develop and integrate the primary services to categorize Technology Evolution (TE) with the help of the Mobile Payment landscape. 
\\ \hline 
 \cite{solat2017security} 
& The author focused on innovations and new attempts or dominant systems to improve the electronic Mobile Payment System (MPS). 
& This survey consists of a review of its dominant system and present cart transactions (CPT). At Cambridge University, several types of research are conducted to designate different attacks against authentication methods in MPS. 
\\ \hline 
 \cite{kim2019can} 
& The research study aims to analyze and examine the effects of consumers' and consumer preferences for MPS and features of the marginal usefulness of biometric and mobile payments.
& Based on the study results, theoretical effects for mobile payment consumer preference and proposed different market strategies for the dispersal of main next-generation MPS from different aspects are analyzed. 
\\ \hline 
 \cite{liebana2019use} 
& The objective of the author is to analyze the factors that affect users' intention in the Mobile Payment System and the status of Near Field Communication in the transportation system.
& To achieve the objectives of the research, a survey was completed with 180 mobile payment users. A widespread review of scientific collected works validates the progress of interactive model that clarifies the intention of near field communication MPS with the help of structural equation model (SEM). The study results show that perceived risk, effort expectancy, service quality, and satisfaction define the persistence aim to use MPS. \\ \hline
 \cite{de2019mobile} 
& The research study compares the different factors that determine consumer acceptance quick response(QR), near field communication (NFC), and short message service (SMS) in MPS. 
& To achieve the objectives, the intention to use mobile payments, a comprehensive review of literature has necessary for the improvement of the behavioral model. The novelty and results of the study lie in the preparation of different behavior rendering to use given by MPS users to each planned payment application. 
\\ \hline 
 \cite{sumathy2017digital} 
& The research study focused on the urban consumer's perception and attitude towards digital MPS. 
& A convenient survey was conducted to achieve the research objective among 100 urban respondents with an interview schedule. 
The ranking method, independent sample t-test, one-way Enova, and percentage analysis are used in this study.
\\ \hline 
 \cite{fatonah2018review} 
& The author Review the E-MPS in E-Commerce. 
& This study aims to analyze the available literature related to e-payment and e-commerce to underline the possibility of e-payment and identify the research gaps, and for future studies, the methodology of previous researchers is recommended.
\\ \hline
\end{tabular}}
\end{table*}

\subsection{Mobile Payment System (MPS): History, Development, and Deployment} \label{MPS-History}
MPS provide several payment facilities for different kind of services, products, and bills through mobile phone by using wireless characteristics and other features and benefits of a communication system \cite{evans2015empirical}. Mobile devices like smartphones, smart tablets are utilized in different payment scenarios such as purchasing online tickets, electronic materials, online electronic transactions, and transport fares such as paying bills and other invoices. It is also possible to purchase products physically through MPS, either from the point-of-sale (PoS), ticketing machines, and vending machine stations. Besides that, most electronic payment systems and payment instruments nowadays have also been mobilized \cite{van2014users}.\\
The field of MPS is relatively new, and little is known about it. Mobile phones are more than just a payment method. Instead, it is a method of initiating, processing, and confirming financial transactions. Mobile payments are not only about using mobile devices to access online payment services. While the mobile version of the service may have similar functionality, the design and implementation of mobile payments are also different due to different methods and structures.
Numerous factors boost MPS evolution in developing countries. Following are some of the factors.
\subsubsection{Socioeconomic Conditions}
The lack of cash alternatives is the most critical factor fostering MPS growth in emerging developing economics countries \cite{evans2015empirical}. Maximum people in developing economies countries have not checked accounts and have not to debut or credit card. Well-developed mobile payment applications with the advantage of low fees for money transfer services from one application to other make MPS attractive \cite{van2014users}. In almost all countries, people move toward the mobile banking system to save their valuable time and avoid getting robbed. 
\subsubsection{Cost Efficiency}
In developing countries, most online conducted transactions are very low in terms of value, but they are very high in volume \cite{dodini2016consumers}. Introducing a new bank branch is infeasible because of massive initial equipment, investment requirement, infrastructure, and well-trained HR included security staff. Bank without different branches looks appealing because it utilizes local infrastructure and leverages local resources and human resources and equipment and resources, including agent shops and mobile phones. Mobile Payment Systems (MPS) are reflected as valuable because of their bottom-of-the-pyramid, lower-class families and unbanked population. The fee for a usual payment transfer is almost 1\% in all mobile payment systems. E.g., the fee for sending money through Wizzit and MTN in South Africa (SA) is almost US\$0.05. But earlier than the Wizzit and MTN payment system, the average fee is almost US\$30 to US\$50 for the delivery of cash. 
\subsubsection{Diffusion of Mobile Phones}
As cell phones become cheaper, financial systems are still relatively limited, and Mobile Payments (MP) are more convenient \cite{duncombe2011researching}. In most countries, individuals may have one or more cell phones. Sub-Saharan Africa has more families with more cell phones than sustainable electricity or drinking water resources.
\subsubsection{Convenience}
In advanced countries, MP is more suitable. People can pay or withdraw money without leaving their homes, which will significantly save their time and cost of the expensive fees. However, this is not an issue in advanced countries, as ATMs and banks are opposite \cite{dermish2011branchless}.
\subsubsection{New Initiatives}
Non-governmental organizations and international organizations (e.g., IFC, the World Bank, GSMA, Gates Foundation) have proposed new initiatives to promote and facilitate MPS implementation. For example, M-PESA Kenya was launched and developed by Safaricom and Vodafone with help from UK's Department for International Development. Pakistan telecommunication (PTC) (Easypaisa) received a \$ 6.5M grant from Gates Foundation in 2012. On the other side, many factors limit the growth of MP.\\

On the other side, several factors are limiting further MP development.

\subsection{Factors Limiting MP Development}

\subsubsection{Heavy Restrictions and Regulations}
This is the most destructive factor in the development of mobile payments. Pressure on banks plays a key part in the ecosystem also decreases the development of MP. Unfortunately, compared to technological advances, most mobile payment methods are changing slowly \cite{van2014users}.
\subsubsection{Limited Collaboration}
In most situations, non-cooperation is an obstacle to the ecosystem. For example, M-PESA has worked with commercial banks for five years to ensure that their valuable customers withdraw their money from ATMs and banks. Collaboration is very significant as most customary banks do not implement to handle MP.
\subsubsection{Underdeveloped Ecosystem}
Lack of standards, undeveloped infrastructure of systems, limited mobile resources, and saturated telecommunications networks (including disruptions) prevent developing countries from launching Mobile Payment Systems (MPS) \cite{liu2015competition}. In some situations, interoperability concerns and a specific type of broker are needed to solve the trust problem and reduce the chicken and egg problem.
\subsubsection{Security Problems}
Cybercriminals' activities are more in advanced countries concerning others countries. First, advanced countries often lack an adequate legal framework and implementation tools to fight cybercrime. Secondly, occasionally customers have not knowledge and attention is very little to security problems. This means that high technology is unlikely to be suitable for developing countries.\\

Given the various influences that drive and delay the development of MP, all critical factors in the ecosystem must be focused on the longstanding goals of the MPS. Of course, the utmost important objective of any MS is to improve competence, conducive to financial development. In MP, it is an alternative to financial transactions and specializes in small payments that cannot be made in cash. However, it remains to be seen whether the key players in the development and implementation of the technology are willing to make large-scale commitments \cite{iman2018mobile}. \\

Given the various influences that drive and delay the development of MP, all critical factors in the ecosystem must be focused on the longstanding goals of the MPS. Of course, the utmost important objective of any MS is to improve competence, conducive to financial development. In MP, it is an alternative to financial transactions and specializes in small payments that cannot be made in cash. However, whether the key players in developing and implementing the technology are willing to make large-scale commitments remains to be seen. MPS lead to the growth of new marketplace ecosystems, containing mobile operators, card operators, retailers, service providers, banks, hardware vendors, trusted service managers, and technology vendors. Several critical regulatory issues emerged, such as electronic money and payment systems, consumer data protection, MPS, principles, and confidentiality. MPS are used in developed countries and Asia, and Africa. Mobile payment systems are used for interpersonal transfers (P2PT), handling small purchases, paying bills and expenses, and purchasing specific goods or services. Almost all mobile network operators that provide mobile payment systems operate in the few countries/regions they are located in, thus facilitating international transactions and remittances \cite{iman2018mobile}.\\

There are no separate laws for MPS in several cases, especially in undeveloped countries. On the other hand, depending on the types of mobile, payment, retail and convergent value chain technologies described and classified above, the program is multifaceted and extensive \cite{iman2018mobile}. The bond structure is unmoving in its beginning but applied in all areas and at all system levels. With the development of technological threats and economic and financial benefits, mobile payment systems began to develop. The regulatory issue of mobile payments is new for at least two reasons. First, it summarizes the different areas of data privacy, e-money, ICT, mobile services, e-payments, user protection, and information and rules and regulation. Second, there are some specific problems with innovation, namely the interpretation of electronic money and the oversight of payment systems.
\subsection{Mobile Payment System Key Attributes}
All mobile payment systems provide greater convenience of using mobile devices to process electronic payments \cite{ozok2010empirical}. However, it should be noted that because they perform many functions in a universal payment system, mobile payment services have different features that will affect the preferences and decisions of the user. Therefore, mobile payment services have complex features, including a combined process of merchant visits, identity verification, and payments. Table \ref{tab:MFS} presents the mobile financial system (MFS) key attribute. \\
All mobile payment systems provide greater convenience of using mobile devices to process electronic payments \cite{ozok2010empirical}. However, it should be noted that because they perform many functions in a universal payment system, mobile payment services have different features that will affect the preferences and decisions of the user. Therefore, mobile payment services have complex features, including a combined process of merchant visits, identity verification, and payments. It is still significant to explain the concept of MFS, containing mobile banking (MB), mobile payments (MP), mobile wallets (MW), and mobile currency (MC). Considerate these facilities are the main research encounter in mobile money transfers \cite{shaikh2015mobile}. MB mentions providing banking services through mobile communication devices, including financial transactions (for example, money orders and bill payments) and non-financial business transactions (for example, balance surveys). Some researchers believe that the functions of MB and MC intersection \cite{slade2013extending}.\\

\begin{figure}[h!]
\centering
\includegraphics[width=\columnwidth]{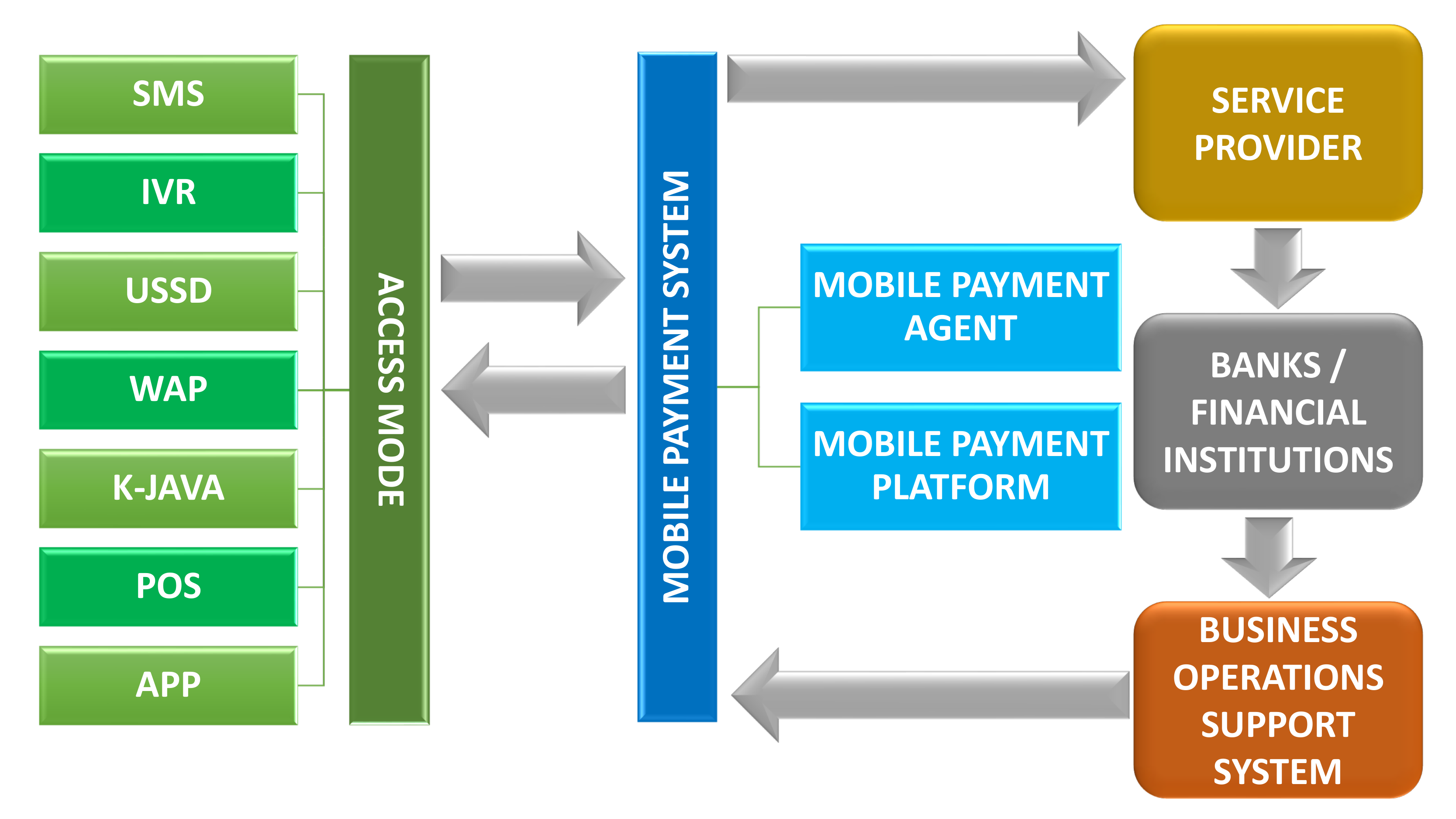}
\caption{Structure Of MPS}
\label{fig:PTyyyy}
\end{figure}

While MB is primarily seen as a straight link between consumers and banks \cite{oliveira2016mobile}, mobile payments are categorized as a service technique that affiliated service suppliers can use deprived of the involvement of banks. Mobile payments are common and generally refer to any payment that uses a mobile terminal to confirm and authorize a payment transaction \cite{kumar2019driving}. Alternatively, mobile wallets are defined as progressive mobile applications that replace physical wallets and have numerous functions like storing payment info and affiliation card and executing transactions. Finally, the mobile currency is a currency that can use and accessed via an MP. Especially since it allows users to run a business (e.g., money orders) without a bank account, it is extensively used among rural inhabitants and cannot use traditional financial institutions \cite{glavee2019drivers}.
\begin{table*}[!h]
\caption{Mobile Financial System}
\label{tab:MFS}
\centering
\begin{tabular}{|l|p{11cm}|}
\hline
\textbf{Mobile Financial System} & \textbf{Description} 
\\ \hline
Mobile Currency 
& MC is a currency that can be opened and used by an MP deprived of a bank account \cite{glavee2019drivers}.
\\ \hline 
Mobile Payment
& MP is a payment system used by an MP to authorize, initiate and confirm a transaction \cite{kumar2019driving}. 
\\ \hline 
Mobile Wallets 
& MW is a mobile application that can replace physical wallets and include the following features: The ability to store information about payments, membership cards, and membership cards, and other marketing plans \cite{kumar2019driving}. 
 \\\hline
Mobile Banking & MB means providing banking services via mobile phones, including financial and non-financial transactions. Wireless device \cite{slade2013extending}. 
\\ \hline
\end{tabular}
\end{table*}

\subsection{M-Payment Stakeholders and Entities} \label{Sec2}
\subsubsection{Stakeholders in MPS}
There are many diverse stakeholders in implementing M-Payment, including consumers/clients, merchants/providers, mobile network operators (MNO), mobile device manufacturers, financial institutions, banks, software, and technology providers. The government is the stakeholder in the M-Payment implementation process. Each stakeholder has different incentives, roles, and strategies. Sometimes these interests and strategies between different stakeholders conflict, e.g., the network provider would like to maximize revenues through each m-payment transaction, whereas customers and merchants would like to minimize costs for each M-Payment transaction. In another study \cite{nejad2016stakeholders}, the author highlights the critical finding that mobile payment method depends on their providers to connect the merchants and consumers to the degree that satisfies the stakeholders.

\subsubsection{Communication Entities in MPS}
\begin{figure}[h!]
\centering
\includegraphics[width=\columnwidth]{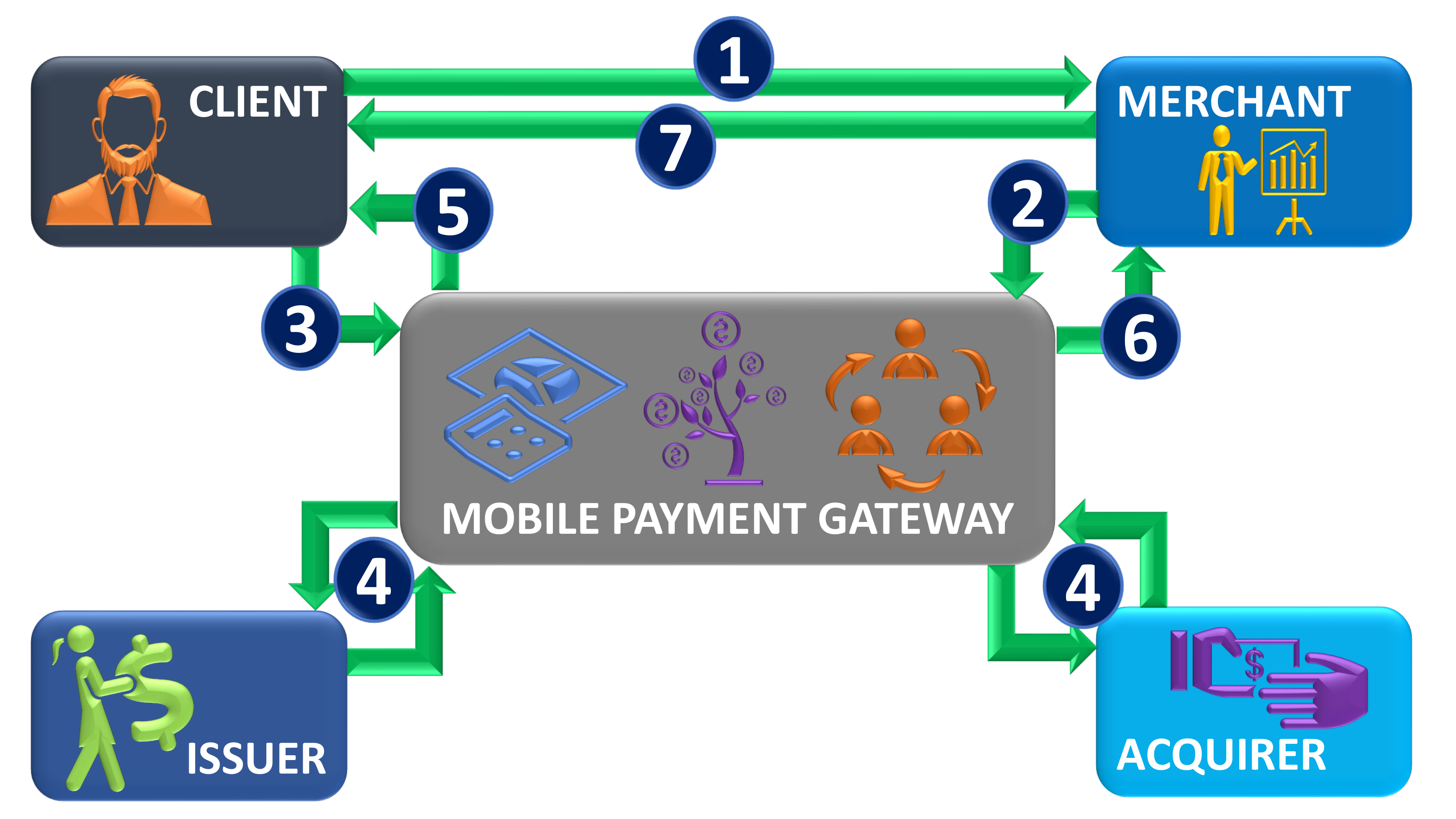}
\caption{Entities involvement in M-Payment process}
\label{fig:EntitiesInvolvementMPaymentProcess}
\end{figure}

For the payment process, there are multiple entities (as shown in Table \ref{tab:1}) that perform their role. In Figure \ref{fig:EntitiesInvolvementMPaymentProcess}, \cite{isaac2014secure} shows the entities that communicate in mobile payment process. The entities can be less or more according to the protocol.
\begin{table*}[h!]
\centering
\caption{Entities that involve during mobile payment process}
\label{tab:1}
\begin{tabular}{|l|p{12cm}|}
\hline
\textbf{Entities} & \textbf{Description} \\ \hline
Client & An entity who wants the transaction 
\\ \hline 
Merchant & An entity that has products or services to sell. It could be a computational one (like a standard web server) or a physical one. 
\\ \hline 
Payment Gateway & another entity acts as an intermediary between the acquirer/issuer on the bank's private network side and the client/merchant on the Internet for payment clearing purposes. 
\\ \hline 
Issuer & The client's financial institution manages the client's account and affords the electronic payment instruments to be used by the client. 
\\ \hline
Acquirer & The merchant's financial institution manages the merchant's account and verifies the deposited payment instrument. 
\\ \hline
\end{tabular}
\end{table*}

\textbf{Steps that involve in M-Payment process}

\begin{enumerate}
\item Client request to a merchant for the payment.
\item Merchant requests to the payment gateway for the transaction amount to be a deposit.
\item Client request to the payment gateway for checking the deduction amount from the account. 
\item Payment clearance is held in the payment gateway. 
\item Payment gateway response to the client request in the form of rejection or approval. 
\item Payment gateway response to the merchant request in the form of acknowledgment receipt.
\item Merchant gives the payment receipt to the client and confirms the transaction.

\end{enumerate}

\begin{figure}[h!]
\centering
\includegraphics[width=\columnwidth]{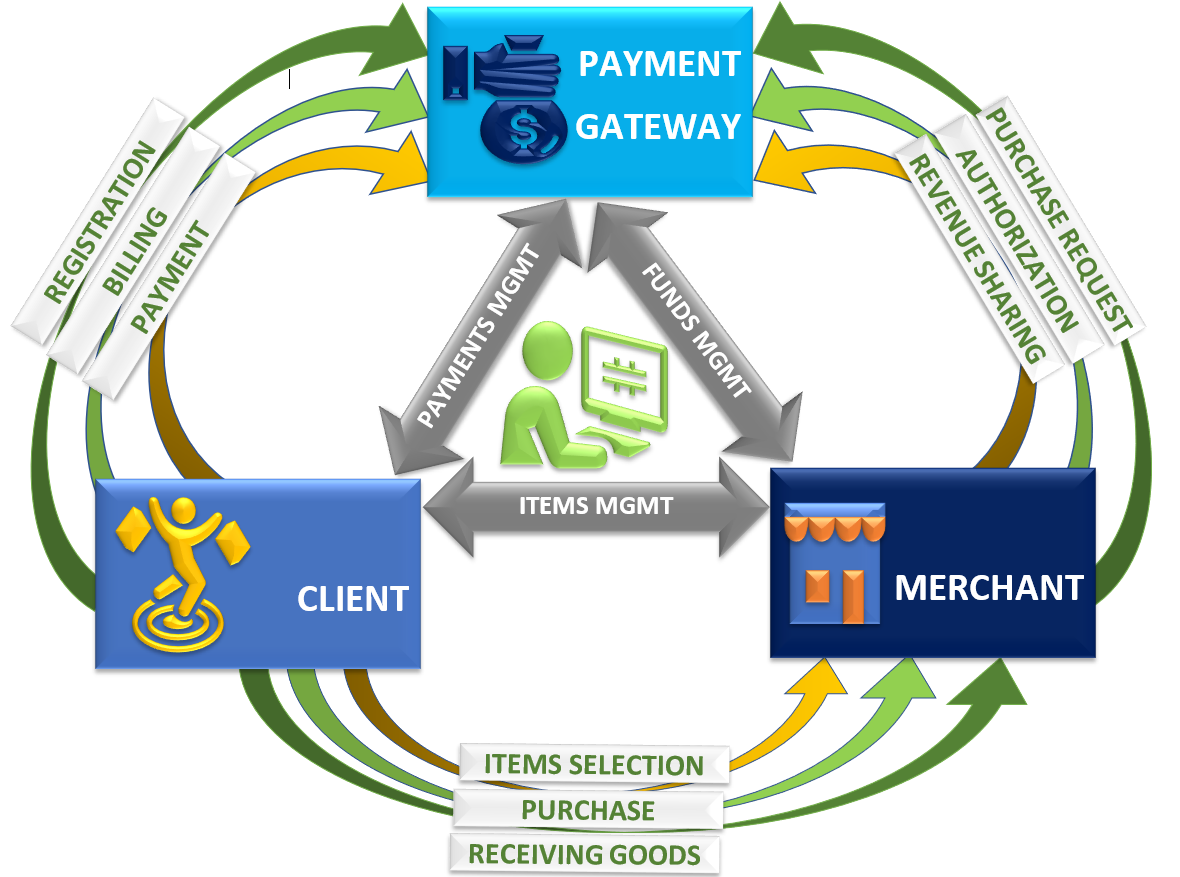}
\caption{Primitive transactions}
\label{fig:PT}
\end{figure}

Figure \ref{fig:PT} represents the model of primitive transactions in which the client makes payment to the merchant. The value of the payment is subtracted from the client's account on the issuer's request by the payment system, and then on the request of the acquirer, the merchant transfers/adds the value from the payment gateway to its account.

\section{Mobile Payment System Security Mechanism}
MPS security mechanism included: Encryption technology, authentication, and a firewall \cite{sun2019mobile}.

\subsection{Encryption Technology}
Encryption technology included: Symmetric encryption and public-key encryption.

\subsubsection{Symmetric Key Encryption (SKE)}
SKE system uses a common key to encrypt messages, which means both sender and receiver will hold a common key for encryption and decryption. Before transmission of data between both parties, the common key is shared on the secure channel between both entities \cite{chaudhury2017acafp}. Exchanging keys between both entities is important for encryption processes. Short size and weak keys are easily attacked opposite to longer keys. Symmetric encryption is still commonly used in insecure data communication. 

\subsubsection{Public-Key Encryption (PKE)}
PKE system is a type of asymmetric encryption because the same key is not used to encrypt and decrypt the messages. In the PKE system, two different keys are used, called public and private key \cite{chaudhury2017acafp}.

\subsubsection{Comparison between SKE and PKE}
There are numerous differences between the SKE system and the PKE system. Table \ref{Comparision} presents the comparison of the SKE and PKE. 

\begin{table*}[!h]
\caption{Comparison of Encryption Methods}
\label{Comparision}
\centering
\begin{tabular}{|p{4cm}| p{5.5cm}|p{6.5cm}|}
\hline
\textbf{Characteristics} & \textbf{Symmetric Key} & \textbf{Public Key}
\\ \hline
Several keys are used for encryption, and decryption & The same key is used for encryption-decryption & Two different keys are used for encryption and decryption.
\\ \hline 
Speed of encryption and decryption & Faster than public-key encryption & Slower than symmetric key encryption 
\\ \hline 
Size of ciphertext & Usually less than or same as the plain text & More than plain text 
\\ \hline 
Key exchange & A big problem & No issue
\\ \hline 
Key usage & Used for confidentiality but not for digital signature & Used of confidentiality and digital signature as well 
\\ \hline
\end{tabular}
\end{table*}

\subsection{Authentication}
Authentication included: Digital signature and certificate authority.

\subsubsection{Digital Signature}
Digital signature (DS) is a string value calculated using text value to a Hash value. DS is used to verify the origin of the received text and prove whether the received text is without any changes. To certify the availability of DS, PKI is frequently used. It suggests a complete set of security assurance and follows different public key encryption standards for different sectors like online banking, e-banking, e-government, and e-commerce securities \cite{zhang2010study}.

\subsubsection{Certificate Authority}
The Certificate Authority (CA) is a trusted organization that publishes and manages network security public keys infrastructure (PKI) and credentials for message encryption. As part of the PKI, the CA will use the registry for verification. Users have the right to verify the information in the digital certificate provided by the applicant. Suppose RA (Register Authorities) verifies the applicant's data and issue a digital certificate. Communicates users are responsible for distributing and revoking certificates. Depending on the PKI, Upon request, the certificate may contain the holder's public key, the certificate, the name of the certificate holder, and other information about the holder of the public key \cite{al2012development}.

\subsection{Firewall}
The firewall can simultaneously protect the system /local network against network-based threats. The firewall allows access to the outside world to the local network. In most scenarios, a firewall is necessary because it is difficult to equip all devices with different security devices. Typically, the firewall is inserted between two networks.

\section{Authentication Methods in MPS}
Authentication methods are widely used to test user identity in mobile transactions as the user identity is required to execute transactions \cite{ogbanufe2018comparing}. Below are some authentication methods: knowledge-based authentication verification, object-based authentication verification, and biometric authentication. With Knowledge-Based, users used personal identification number (PIN) or password to validate their identity \cite{jain2007handbook}. This is based on well-known traditional authentication methods, so they have fewer security issues. Use physical tokens (such as smart cards) to perform object-based authentication. While objective knowledge-based methods can create inexpensive and straightforward authentication systems for various computing applications, they can close the security incident vulnerabilities. In addition, the above two methods are likely to be lost or forgotten by the users, which can be an intellectual burden for application users \cite{weir2010usable}.\\
To overwhelmed the restrictions of the above traditional methods, some advanced authentication systems have been developed that provide consumers with helpful security \cite{crawford2013framework}. Biometric methods based on user personal identity (i.e., Physical Characteristics) have been effectively applied to protect and verify users' identities. Identity verification based on human-specific biometrics (such as fingerprints, voice, or iris) is unlikely to be easily stolen or transferred. Previous research has shown that the apparent advantages of biometric authentication systems improved account security and the perceived pleasure and reduced cognitive load and time, which provides more excellent value to consumers \cite{byun2013exploring}.\\
Biometric technology has its limitations. Because of the complication of high-quality images, many factors in a biometric system will reduce user identification accuracy \cite{alsaadi2015physiological}. Physical issues like wet surfaces, dirty fingers, or scratches are familiar illustrations that can delay biometric authentication. In addition, the biometric system also has some privacy issues related to users' identity management. However, there is also an advanced "Knowledge Based" authentication method. In the meaning of some useful security features, graphical prompts (such as design drawings) have also been proposed as a substitute to the above authentication methods \cite{alzubaidi2016authentication}. A recent study found that using different e-payment authentication methods will affect users' perception of security and availability of these three authentication types. Therefore, in this study, the method of identity verification was selected as one of the primary resources for creating the preference structure of mobile payment users \cite{ogbanufe2018comparing}.

\subsection{Types of Authentication Factors}
Three types of authentication factors named single-factor authentication (SFA), two-factor authentication (2FA), and multi-factor authentication (MFA) can be understood through the definitions proposed by the research of \cite{SFA123} and \cite{Tssw123}. They proposed that a process allowing individual users to seek access from authenticating parties for attestation of their personalities with the utilization of single attribute associates with their identifies is termed as Single-Factor Authentication (SFA). An example of such an attribute would be the use of a PIN for unlocking cell phones. The user-friendly and straightforward nature \cite{ometov2018multi} of this authentication type made it a preferable choice for many companies; however, its vulnerability to various forms of attacks \cite{Tssw123} made it unsuitable for application in financial institutes. \cite{Tssw123} Defined Two-Factor Authentication (2FA), mentioning that users seeking requests for access from authentication party through attestation of their personality with two attributes are a process that comes under the concept of Two-Factor Authentication. These attributes include knowing something personal or possessing something personal that can be associated with one's personality. Hence, attackers are bound to be aware of two identifiers to get the same authentication as the original users in 2FA. This particular feature of 2FA makes it acceptable and applicable by financial institutions. However, this type remains loopholes, leaving it vulnerable to a MITM attack, eavesdropping, and Trojan horse attack. Furthermore, it has its limitations when considered for its effectiveness against phishing \cite{ometov2018multi}.\\
For defining the third type of authentication factor named Multifactor Authentication (MFA), \cite{Tssw123} describes that it involves users seeking requests for access from authentication parties through attestation of their personality with multiple attributes make up. Biometrics are used along with ownership and knowledge as an attribute by MFA. MFA's higher level of security makes it a better choice for various critical services and computing devices. Physical separation of authentication factors from the user device can allow MFA to be more successful. The addition of biometric factors makes MFA achieve improved identity proof resulting in more secure systems \cite{ometov2018multi,hamilton2017database}.

\section{Cyberattacks on Mobile Payment System}
Different level attacks on MPS can come from unauthorized malicious users. Following are some identified attack points susceptible to comprise in this regard.

The first attack is targeted at the users of mobile money. It includes accessing the PIN of users via shoulder-surfing when it is unmasked PIN of four to five digits \cite{lakshmi2017ussd}. Access to this PIN can enable attackers to make fraudulent transactions. Brute force attacks can also be performed by attackers considering the straightforwardness of the PIN \cite{castle2016let, reaves2017mo, mahajan2015mitigating}

The second type of attack involves comprising of money communication channels. The hacking and controlling of MMS traffic and manipulation of accounts for making transactions can be made possible using these points \cite{castle2016let, reaves2017mo, mahajan2015mitigating}.

The third type of attack is at the server of the mobile money app. Availability of server to both mobile money agents and users is suspended when such attack is carried out at server. As per the findings of Castle et al. \cite{castle2016let}, attackers divert fake traffic to mobile money servers resulting in it being overwhelmed, which eventually leads to blocked requests from mobile money agents and users. It can also include installing malware on the mobile money app server for deducting some amount from wallets of mobile money agents and users for deposition into the attacker's account without letting these users or agents discover the transaction \cite{salahdine2019social}. 

The fourth point of attachment is the IT administrator. The administrator's computer can be hacked by an unauthorized person making it inaccessible to the administrator by changing its credentials.
Mobile money agents can be considered as another attack point. The PIN of the commission agent can be stolen by an attacker using shoulder surfing techniques. Attackers can also practice giving the wrong PIN repeatedly while making transactions to access agents' PINs. \cite{buku2017fraud} and \cite{lonie2017fraud} Identified adversaries gave that wrong phone numbers repeatedly to obtain the PIN of agents and use it for gaining unauthorized access to the float accounts of agents.

Bank's server provides another attack point for adversaries. A distributed denial-of-service attack (DDoS) is made in such cases to create the unavailability of a bank server to the mobile money user trying to make a transaction.

Notification message channels where messages can be modified creates another attack point for malicious users. Adversaries may hack the communications channels of the notification message and make changes in the message as per their requirements while sending the modified versions of these messages to the intended users \cite{ sharma2019contemplate, kumari2020eseap}.

\subsection{Attacks against Privacy}
 \cite{katusiime2021mobile} Defined privacy as the right of users to have freedom from intrusions and infringements by other users. In mobile money, privacy attacks include the compromised PINs of the users for illegal access to their financial assets and information details utilized in unauthorized transactions. Stealing of user information can result in a problematic situation for not only the user but also for the economy as well \cite{katusiime2021mobile}. Illegal access to the mobile money database containing the financial information of users can allow attackers to update or delete records using the stolen PINs.\\
Moreover, a variety of user-related information can be stolen when an attacker gets access to mobile money database \cite{katusiime2021mobile}. Personal information such as email addresses, mobile telephone numbers, NIN, and even names of users and agents can be compromised, failing privacy safeguards \cite{katusiime2021mobile}. Unscrupulous insiders may end up abusing highly sensitive data after gaining control and access in this way. Attackers can do so with the generation of a databank to give control and access to personal information. There are situations in which some users request the agents for assistance in performing transactions, and they end up sharing their PINs with the agents \cite{mckee2015doing}. It raises the bar for the required level of protection to agents and mobile money users against unauthorized access.

\subsection{Attacks against Authentication }
The identity of a user is forged by an attacker impersonating an authorized user in this form of attack. According to \cite{ali2020evaluation} authentication attack is a crime in which the mobile money authentication process is subjected to exploitation when a brute force attack is being carried out against the PIN. Various attacks are included in this form of attacks, such as Trojan horse attack, phishing attack, social engineering attack, spoofing attacks, masquerade attack, replay attacks, and impersonation attack. An attacker assumes the identity of a legitimate user in an impersonation attack \cite{ali2020evaluation, gwahula2016risks, buku2017fraud, lonie2017fraud}, whereas entire communication is subjected to eavesdropping in replay attack before intercepting \cite{ali2020two}. In a masquerade attack, the PIN and SIM card are acquired by the users.\\
Moreover, an attacker pretends to be a mobile system administrator in a spoofing attack. When users are manipulated for them to give up their personal information, a social engineering attack is said to be launched \cite{kunda2018survey}. Similarly, a phishing attack involves deceitful attempts by adversaries for accessing personal information needed to impersonate a legitimate user in the system \cite{ali2020evaluation}. Another method of compromising an authentication system involves using Trojan software as a virus to access users' personal information.

\subsection{Attacks against Confidentiality}
Attacks on confidentiality involve eavesdropping on the communication channels between the application server and mobile money users for tapping information like PIN of users that can be used for impersonation or in making unauthorized transactions. The four types of such attacks include guessing attacks, brute force attacks, eavesdropping, and should surfing attacks. Attackers secretly hear the communication channels in the eavesdropping attacks by taking advantage of the lack of security of the network communication. The plain text form of transmitted data is commonly vulnerable to such attacks \cite{talom2019impact, nyamtiga2013enhanced}.\\
For brute force attacks against confidentiality, the adversary can guess the mobile money agent or user's PIN needed to access the mobile money account. Despite being very simple, such types of attacks have shown a high rate of success \cite{ali2020evaluation, gwahula2016risks, buku2017fraud, lonie2017fraud}. When an adversary sees mobile money PIN during authentication, a guessing attack is being launched. Shoulder surfing attack also comes under the type of attacks against confidentiality in which the adversaries acquire confidential data and PINs simply by looking over the shoulder of the victim as they make transactions \cite{ gwahula2016risks}. \cite{9432677} discussed one of the approaches that can be utilized to extract useful network information to identify the details of the endpoint of the communication parties on the local network.

\subsection{Attacks against Integrity}
When information of the user is accessed and modified in the MMS, the integrity of user information is compromised. They can be categorized into insider attacks, salami attacks, and MITM attacks. An intruder intercepting the communication between various agents (including users) in the mobile money application network performs Man a middle attack. Sitting between the mobile money user and MMS, the attacker makes them believe they communicate in the MITM attack. \cite{nyamtiga2013enhanced} and \cite{mahajan2015mitigating} Showed that an attacker could gain control over the entire conversations when a MITM attack is being launched as the content of the conversation is modified by the attacker at both ends.\\
Employees of financial institutions can conduct salami attacks and insider attacks against the financial institutions. Like a Trojan horse, a salami attack involves installing malicious software to a financial institution system, allowing adversaries to withdraw money from users' accounts, depositing it in their accounts. Both external and internal adversaries can launch salami attacks in which small deductions are made to user wallets as the software allows modification of some details in the system \cite{chowdhury2016recent, sadekin2016security, altwairqi2019four}. \cite{shaw2014mediating} and \cite{kshetri2019cybercrime} Highlighted the fact that a high degree of risk of money fraud comes from the employees of the financial institutions who are aware of the security protocols and MMS in the system. MMSP employees with sufficient knowledge about the organization's security practices can be involved in the insider attacks identified by \cite{ mckee2015doing}, and \cite{ shaw2014mediating}.

\subsection{Attacks against Availability }
When the bank server or application server is suspended for agents and mobile money users by an adversary on purpose, an attack is assumed against availability. Services are rendered unavailable by adversaries in this type of attack using various techniques. Mobile theft, DOS, and distributed denial of service (DDoS) attacks come under this category. DDoS and DoS are launched when adversaries send fake traffic to overwhelm servers to block legitimate traffic or requests of users \cite{ ali2020evaluation}. Such attacks aim to flood mobile money servers with so many fake requests that the server fails to receive and respond to legitimate users' actual requests, making the service unavailable.\\
 \cite{ ali2020evaluation, reaves2017mo} Described mobile phone theft attacks as the type of availability attack in which the mobile phone of users or agents is stolen, and the wallet account of SIM card is made unavailable that can be swapped by the attacker. Service and data access can be lost due to phone theft attacks, as the attacker can take charge of the victim's e-wallet account, resulting in its unavailability to the actual user.

\section{Technologies Used in M-Payment Process}\label{Sec3}
\begin{figure*}[h!]
\centering
\includegraphics[width=\textwidth,height=08cm]{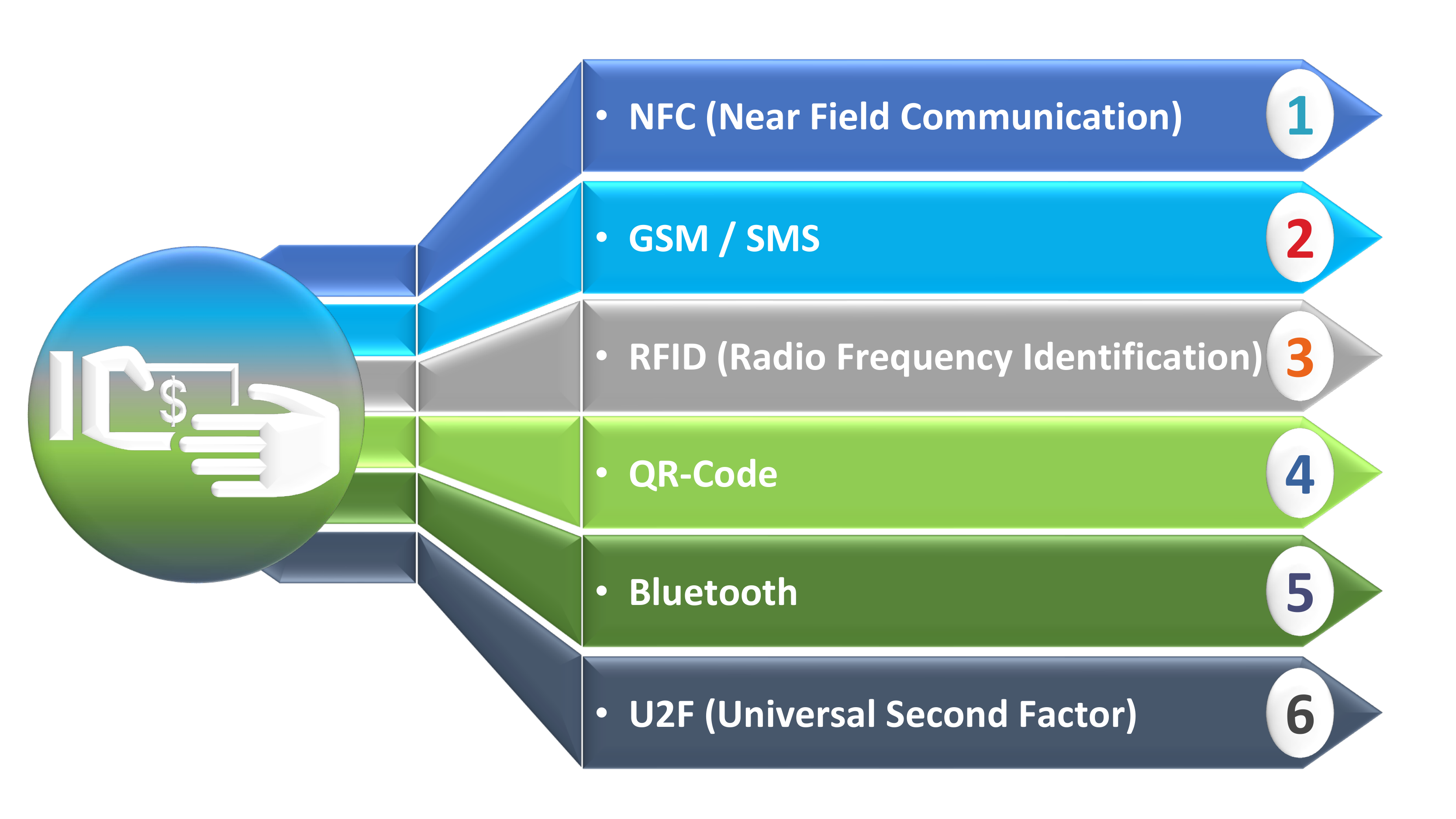}
\caption{Technologies used in M payment system}
\label{fig:TechnologiesUsedinMpaymentSystem}
\end{figure*}

We have also seen an M-Payment system based on the technology used; it is classified in Figure \ref{fig:TechnologiesUsedinMpaymentSystem}. M-Payment system uses mobile technology for communication between the entities involved in the payment process.

Near field communication (NFC) \cite{tamazirt2017nfc} is a communication protocol that enables the communication between two devices. Global system for mobile (GSM) \cite{bhatta2017gsm} is a standard system for mobile communication. Radiofrequency identification (RFID) \cite{tsao2016supply} uses an electromagnetic field to identify or track tags attached to an object. Short messaging service (SMS) \cite{dix2017investigating} is a text messaging service that is used for communication over the mobile phone. Quick Response Code (QR-Code) \cite{mukherjee2017scheme} is a two-dimensional matrix barcode, which has a label in which information is stored. Bluetooth \cite{hassan2018security} is a standard for wireless technology; by using this, we can communicate to fix devices over a short period of distance. Identity-based signature (IB-Signature) \cite{deng2017identity} is a type of public key infrastructure (PKI) in which a publicly known string that represents an individual is used as a public key, e.g., email address, the wireless application protocol (WAP) \cite{kizza2017security} is a standard protocol used in the wireless network to access information. Universal 2nd Factor (U2F): It is a standard of open authentication which provides two-factor secure authentication.

\subsubsection{Reviewed Approaches}
This study has reviewed multiple schemes or models of mobile payment systems based on different technologies or architectures. Paper \cite{qadeer2009novel} uses RFID \& SIM, which enables users to use their mobile phone as ATM card/credit card. By using RFID readers at ATM, users can withdraw money. Paper \cite{manvi2009secure} uses Bluetooth with Java technology by which users can pay at POS (point of sale) by using a mobile phone. Java components are used here to provide encryption. Paper \cite{liu2005system} works on existing models; it enhances SEMOPS (Secure Mobile Payment Service) model by involving a trusted third party.
Papers \cite{zheng2003study, chen2010nfc, al2017online} used GSM technology to implement a secure m-payment system. It provides low-cost architecture by using the existing GSM mechanism. Papers \cite{chen2010nfc, al2017online,nseir2013secure, chen2016nfc} used NFC communication, which provides more speed for communication than other technologies. In \cite{al2017online}, the proposed scheme also provides user anonymity and the un-linkable transaction to defend against attacks. \cite{nseir2013secure, purnomo2016mutual, ma2015design} use QR-Code, which is fast and supports the buy-and-sale process easily and efficiently. In \cite{chen2016nfc} using SMS service; mostly SMS service in mobile communication is used for authentication. In \cite{rui2015design}, using the SMS based authentication system and proposed the application based system in which authentication code can only be accessed by an authorized user and using IB-Signature, which is simple and less costly, the identity of an entity is used in this technology for authentication or for granting access. Authors in \cite{rui2015design} also use the OTP code for securing communication; it prevents the system from replay attack and uses a password for only one time. WAP or Bluetooth technology is used in \cite{harb2008securesmspay} which provides fast communication but over a small range of areas. It is for peer-to-peer communication and less in cost. \cite{firoz2017defensive} uses SMS for sending a notification, but for transactions purpose, it uses unstructured supplementary service data (USSD), which provides a more responsive service than SMS. In \cite{fan2017u2f}, the author uses U2F technology, which is fast and much secure; it performs cryptographic functions with a single touch \& generates asymmetric keys. It authenticates both the client and server in a reliable manner so transactions can be done securely. Query processing over encrypted data is the solution to tackle the extra overhead caused by the use of encryption, and such techniques result in remarkable improvements in the scenario where near real-time data processing is required \cite{7454484}.

\begin{table*}[h!]
\caption{Comparison of reviewed papers on the basis of payment system technology, architecture, communication entity involvement, assumption, advantages and disadvantages of solutions (I)}
\label{comp}
\centering
\scalebox{0.8}{
\begin{tabular}{|p{.8cm}| p{1.8cm}| p{2cm}| p{2cm}| p{2.5cm}| p{2.5cm}| p{3cm}| p{3cm}|}

\hline
\textbf{Ref} & \textbf{M-Payment Based On} & \textbf{Architecture Used} & \textbf{M-Payment Between} & \textbf{Provides} & \textbf{Assumption} & \textbf{Advantage} & \textbf{Disadvantage} 

\\ \hline

 \cite{ma2015design} & RFID & RFID \& SIM & 
User \& POS User \& Service provider User can also deposit money & Easy to pay service and ATM/Credit card functionality through mobile & RFID readers are installed at stations, shopping malls, \& at ATM, & Flexibility, time, workforce reduction, safety \& mobility & Some issues in security \& privacy in RFID Chances of unauthorized use are present in case of theft or loss of mobile.

\\ \hline
 \cite{harb2008securesmspay} & J2EE and J2ME capabilities & Java, Bluetooth & Payment server \& mobile clients & Confidentiality and Authentication & MIDP application is uploaded to client’s mobile & It overcomes the API and technical limitations, as well as security consideration & As the number of clients increases the delay (milliseconds) will increase over Bluetooth environment 

\\\hline
 \cite{ruan2014desgn} & Existing models & SEMOPS \& trusted third elements & Customer \& Merchant & Privacy \& Non-Repudiation & All parties possess  certificates among each other & It introduces trusted third elements \& follows new Mechanism to achieve privacy \& non-repudiation & Difficult to implement

\\\hline
 \cite{singh2012comparative} & Multiple layers & SMS, GSM & Consumer \& content provider & Security \& High Scalability & - & It provides low cost and technical requirement, high scalability, and security & The system required to be more simplified, improve the security \& application of digital signatures 

\\\hline 
 \cite{fan2017u2f} & Scenario for m-payment models & NFC, QR code & Customer \& Merchant & Speed \& Security & - & For speed, transaction is initiated by merchant because he has more reliable \& continuous connection with 3rd party & For every new purchase, there will be authentication by merchant's involvement which can make him busy and it can affect his availability 

\\ \hline
\end{tabular}}
\end{table*}

\begin{table*}[h!]
\caption{Comparison of reviewed papers based on payment system technology, architecture, communication entity involvement, assumption, advantages and disadvantages of solutions (II)}
\centering
\scalebox{0.7}{
\begin{tabular}{|p{.8cm}| p{1.8cm}| p{2.5cm}| p{2cm}| p{3cm}| p{3cm}| p{3cm}| p{3cm}|}
\hline
\textbf{Ref} & \textbf{M-Payment Based On} & \textbf{Architecture Used} & \textbf{M-Payment Between} & \textbf{Provides} & \textbf{Assumption} & \textbf{Advantage} & \textbf{Disadvantage} 

\\ \hline

\cite{rui2015design} & Identity Based Cryptography (IBC) & IB Signatures and One Time Key & Consumer and Merchant & Privacy and Security of transferred data & - & IBC framework is simple and less costly & Higher number of cryptographic operations

\\ \hline

 \cite{chen2010nfc} & GSM & NFC and GSM & Point-of-Sale (POS) and the customer & Security for low value payments, customer anonymity and ubiquitous implementation & Secure channel between payment gateway and shop POS & Re-using existing GSM security mechanisms Payment is same as paying by debit or credit card. & -Short length of encryption key.  -Merchants need to register themselves with mobile operator -protocol is complex as compared to m-payment via SMS or WAP

\\ \hline

 \cite{bojjagani2017secure} & SMS & SMS, WAP or Bluetooth and J2ME. & Payer's or Payee's bank & Secure M-Payment for macro transactions -less encryption or decryption operations & Trusted payment gateway is involved between payer’s and payee’s banks & This scheme shares payee’s financial data with banks only & If mobile got stolen and PIN got also leaked then there are chances of financial loss

\\ \hline

 \cite{kang2017privacy} & PKI & QR-Code and PKI & Client and merchant & Additional layer of security for m-payment systems & Trusted third party is involved to secure encryption keys and to ensure legitimate users & This scheme uses RSA which is considered as strongest asymmetric encryption system & For stronger security longer key pair is required which leads to larger size of QR code

\\ \hline

 \cite{khan2016modified} & 3G & 3G, SMS and IVR & Client and server (Mobile Payment Platform) & Intelligent travel design by using m-payment system & - & By using this user can pay fines, insurance amount or can query traffic violations rules etc & Initial connection in this system takes longer time

\\ \hline

 \cite{luo2016unlinkable} & 2D Bar Code QR 2D Bar Code & Point-of-Sale (POS) and the customer & Advantages to support buy-and-sale products and services base on 2D Barcodes & Trusted third party authentication server is used as Certification authority & - & Products can be traded anywhere, anytime Easy to use Reduce user input & Computations are little complex 

\\ \hline

 \cite{platform2011trusted} & SMS & SMS & P2P Peer-to-peer & It provides features like security, privacy, speed and less cost & - & Money transfer can be done by transmitting mobile number only & Huge number of different device OS and development environment may prevent Support for all devices

\\ \hline

\end{tabular}}
\end{table*}

\begin{table*}[h!]
\caption{Comparison of reviewed papers on the basis of payment system technology, architecture, communication entity involvement, assumption, advantages and disadvantages of solutions (III)}
\centering
\scalebox{0.65}{
\begin{tabular}{|p{1cm}| p{2.3cm}| p{2.5cm}| p{2cm}| p{3.5cm}| p{3.5cm}| p{3.5cm}| p{3.5cm}|}
\hline
\textbf{Ref} & \textbf{M-Payment Based On} & \textbf{Architecture Used} & \textbf{M-Payment Between} & \textbf{Provides} & \textbf{Assumption} & \textbf{Advantage} & \textbf{Disadvantage} 

\\ \hline

 \cite{sureshkumar2017lightweight} & Cryptographic Protocol Shape Analyzer (CPSA) & CPSA & Merchant and payment gateway & Payment mechanism ensuring accountability and un-linkable anonymity less number of cryptographic operations & Customer browses through the merchant website & Customer has an option of making payment using cards of different banks & Limited to a maximum of two gateways 

\\ \hline
 \cite{kang2017privacy} & Traceable signatures, identity-based signatures, anonymous signatures & Signatures & Ternminal and Passenger & Provides mechanism to protect passenger’s privacy & Off-board terminal trusts onboard terminal & It can be applied to off-line mobile payment systems & - 

\\ \hline

 \cite{firoz2017defensive} & Unstructured Supplementary Service Data (USSD) & USSD and SMS & User to Agent and Agent to User & Secured transaction by using two layers of authentication & Mobile wallet number is same as mobile number and there is large no. of agent points across the country & Mitigate human error and prevent cyber-frauds & Doesn’t consider confidentiality and integrity aspect

\\ \hline

 \cite{fan2017u2f} & Universal 2nd Factor (U2F) & U2F, USIM & User to Server & Secure mutual authentication protocol for m-payment systems & - & Provides a reliable service and protect user’s account information and privacy & In registration process, it will take time since it is using asymmetric cryptosystem 

\\ \hline

 \cite{rui2015design} & NFC & Wi-Fi, 4G, GSM NFC & Bank to bank (POS) & Enhance the security of the EMV(Europay, MasterCard, and Visa) exchanged messages & Mobile network operator (MNO) is trusted by NFC enabled Mobile & Adds a security layer to EMV and ensures confidentiality and mutual authentication & It will be failed if POS entity is dishonest 

\\ \hline

 \cite{sureshkumar2017lightweight} & NFC & NFC & User and TSM (Trusted Service Manager) & Secure protocol which is compatible with EMV & TSM and Bank own their key pairs of a PKI cryptosystem & User can perform transaction without disclosing his identity. & Requires a high computation power for TSM and bank 

\\ \hline

 \cite{luo2016unlinkable} & NFC enable phone & NFC and trusted third party & User and Merchant & User anonymity & - & Un-linkable anonymity to user & - 

\\\hline 

 \cite{bojjagani2015ssmbp} & SMS & - & User and bank, bank and gateway & Secure transaction with formal technique & - & Less time take for key generation and encryption decryption, and scheme security is verified by tools & Only for Android and Java 2 Micro Edition device 

\\\hline

 \cite{khan2016modified} & Application Based & Mobile transaction authentication number system & User and Bank & Application-based system that is comparatively more secure than SMS based system & Attacker can get access to web and SMS at the same time & More security than SMS based MTAN & Less efficient than MTAN

\\ \hline
\end{tabular}}
\end{table*}

\subsubsection{Analysis}
The overall average of technological usage of M-Payment systems is depicted in Table 3. We analyzed technologies and found that \textbf{SMS is used in primarily m-payment systems}. The use of NFC, GSM, and QR-Code is also every day in payment schemes or models. They provide many advantages like we know NFC provides faster speed, GSM provides many already implemented services that make these models implementation easy and straightforward, QR-Code is a less costly and straightforward technology used in many mobile payment systems. U2F is much secure and reliable than QR-code, NFC, or Bluetooth.

\begin{table}[h!]
\centering
\caption{MPS technology specific categorization}
\begin{tabular}{|l|l|l|}
\hline
\textbf{Technology} &\textbf{Papers} & \textbf{\%} 
\\ \hline
\textbf{NFC} & \cite{nseir2013secure,chen2016nfc,al2017online,chen2010nfc,rui2015design,luo2016unlinkable} & 28.57 
\\ \hline 
\textbf{GSM} & \cite{qadeer2009novel,zheng2003study,chen2010nfc,al2017online} & 19.04 
\\ \hline 
\textbf{SMS} & \cite{zheng2003study,chen2010nfc,harb2008securesmspay,ruan2014desgn,singh2012comparative,firoz2017defensive,bojjagani2017secure,khan2016modified} & 38.09 
\\ \hline 
\textbf{RFID}& \cite{qadeer2009novel} & 4.76 
\\ \hline 
\textbf{QR-Code}& \cite{nseir2013secure,purnomo2016mutual,ma2015design} & 14.28 
\\ \hline 
\textbf{Bluetooth} & \cite{manvi2009secure,harb2008securesmspay} & 9.52
\\ \hline 
\textbf{U2F} & \cite{fan2017u2f} & 4.76 
\\ \hline
\end{tabular}
\end{table}

\section{Security Analysis of M-Payment Systems}\label{Sec4}
\begin{figure*}[h!]
\centering
\includegraphics[width=\textwidth]{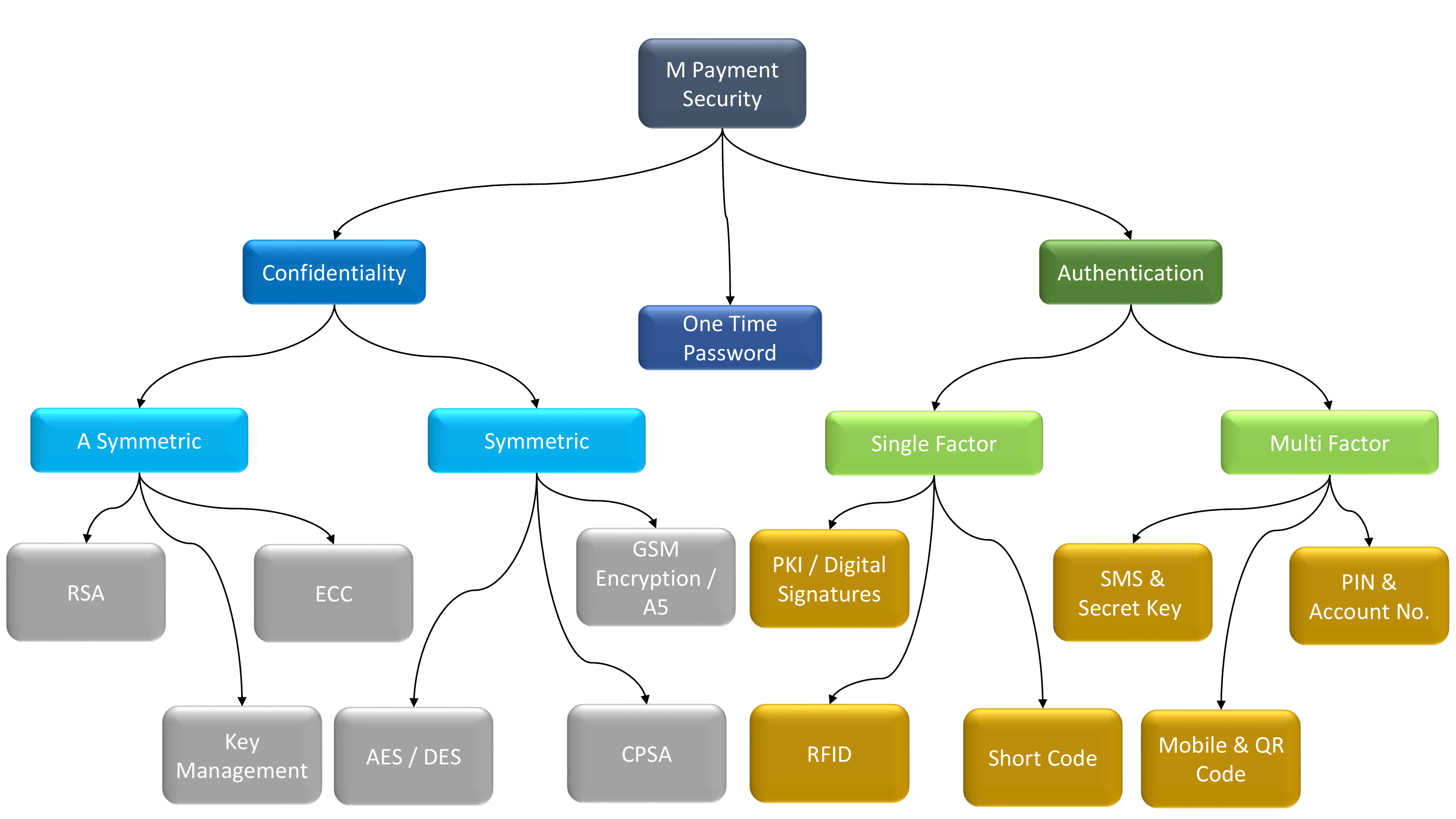}
\caption{M-Payment Security}
\label{fig:MPaymentSecurityServices}
\end{figure*}

This section presents the security analysis of M-Payments system. Security analysis comprise of various services: Authentication, Mutual Authentication, Integrity, Customer Anonymity and Non Repudiation. Figure \ref{fig:MPaymentSecurityServices} the overall service hierarchy of m-payment system. Below we explain each services in detail

\subsubsection{Confidentiality}
In \cite{manvi2009secure}, confidentiality is provided by using Java components. In \cite{liu2005system,khan2016modified} uses cryptography to provide confidentiality. In \cite{rui2015design} OTP and PKI infrastructure are used to provide confidentiality. In \cite{chen2010nfc} GSM security mechanism is used which provide confidentiality via A5 and A8 algorithm. In \cite{chen2016nfc, harb2008securesmspay} confidentiality is achieved by using Symmetric key cryptography to provide confidentiality. Paper \cite{purnomo2016mutual} uses RSA encryption mechanism to provide confidentiality. In \cite{ruan2014desgn} DES and ECC are used to achieve confidentiality. In \cite{ma2015design} confidentiality is achieved by using AES and RSA. In \cite{singh2012comparative} secured end-to-end encryption is used to provide confidentiality. In \cite{sureshkumar2017lightweight} symmetric key encryption is used to provide confidentiality. In \cite{kang2017privacy} RSA encryption mechanism is used to achieve the confidentiality. In \cite{fan2017u2f} Asymmetric cryptosystem is used to provide encryption. In \cite{al2017online} AES is used to provide confidentiality which is a type of symmetric key cryptography. \cite{luo2016unlinkable} provides confidentiality by encrypting the information using asymmetric keys and the key pair stored in the secure storage \cite{platform2011trusted} to protect from unauthorized access. In \cite{bojjagani2017secure} confidentiality is achieved by using ECC which is a type of asymmetric key cryptography. 

\subsubsection{Authentication}
In \cite{qadeer2009novel} authentication is performed by reading the RFID tag, which is embedded in the SIM card. RFID reader authenticates the user in this scheme. In \cite{manvi2009secure} authentication is provided by asking for a PIN and account number. In \cite{zheng2003study} Control and communicating interface are used to provide authentication. In \cite{nseir2013secure} authentication is done by using NFC enabled mobile phone and QR-code/PIN. In \cite{chen2010nfc} triple authentication mechanism is ensured by using the challenge-response protocol. In \cite{harb2008securesmspay} authentication is provided by using SMS and secret key. Paper \cite{ruan2014desgn} ensures authentication by using signature schemes (DES and ECC). Paper \cite{singh2012comparative} uses short-codes to provide authentication. In \cite{sureshkumar2017lightweight} Isomorphic shapes are used for authentication.

\subsubsection{Mutual Authentication}
In \cite{liu2005system,rui2015design,bojjagani2017secure} Payment requests are signed by signing key (Digital Signatures) of both client \& merchant to achieve mutual authentication. In \cite{purnomo2016mutual} mutual authentication is ensured by using the RSA-PKI mechanism. In \cite{ma2015design} RSA-Digital Signature is used to provide mutual authentication. In \cite{kang2017privacy} identity-based signatures are used to achieve mutual authentication. In \cite{firoz2017defensive} mutual authentication is achieved by using Mobile Wallet number and PIN. In \cite{fan2017u2f} mutual authentication is provided by using Asymmetric keys, valid username \& password. For authentication \cite{al2017online} uses Session-key and challenge-response authentication. To ensure mutual \textbf{authentication} \cite{chen2016nfc} uses secret key and public key infrastructure (PKI). In \cite{luo2016unlinkable} authentication is provided by using digital signatures. In this scheme, mutual authentication is only between the user and the bank. 

\subsubsection{Integrity}
To ensure the data has not tampered during transaction \cite{kang2017privacy,bojjagani2017secure,rui2015design,fan2017u2f,chen2010nfc,al2017online} use hash packets and verify the hash. Paper \cite{harb2008securesmspay} uses a private banking network and secure payer confirmation to ensure integrity. In \cite{purnomo2016mutual} integrity is ensured by using QR-Code. In \cite{ma2015design} RSA-Digital signature algorithm is used to ensure integrity. \cite{chen2010nfc} Achieve the integrity by Message Authentication Code (MAC) that is embedded in the ciphertext. In \cite{luo2016unlinkable}, the information is encrypted with shared key among bank/user and signed with user's private key that protects the information from unauthorized modification.

\subsubsection{Customer Anonymity}
In \cite{liu2005system} there is no need to get registered to the merchant or any 3rd party before or during the transaction, which ensures the anonymity of the client. In \cite{chen2010nfc} the client's long-term ID is not revealed to the merchant, which ensures the client's anonymity. In \cite{singh2012comparative} anonymity of consumer is ensured because it only requires the consumer's mobile number or short-code provided by them-payment application service provider. In \cite{sureshkumar2017lightweight} customer's identity is dynamic and updated frequently to ensure the anonymity of a customer. In \cite{kang2017privacy} client's anonymity is ensured by hiding session and transit information. \cite{luo2016unlinkable,chen2016nfc} achieve anonymity by using virtual accounts for clients whom the bank assigns. \cite{lavanya2021novel} proposes a novel approach of ensuring privacy, confidentiality, and authentication using a hybrid scheme for location and payment authentication.
\subsubsection{Non Repudiation}
In \cite{luo2016unlinkable} Non-repudiation is ensured by using the signing key and timestamp. In \cite{rui2015design} IBC-Signatures are used which ensures non-repudiation in their scheme. In \cite{harb2008securesmspay} three factors are used to prove non-repudiation of the client (by checking the status response of the client, session key, and the offline PIN). Paper \cite{ma2015design} uses RSA-Digital Signature to sign transaction information which ensures non-repudiation of transactions. Papers \cite{fan2017u2f,bojjagani2017secure} use signatures to ensure the legitimate user and non-repudiation. \cite{al2017online} ensures the non-repudiation by hashing the transaction data with the shared key. In \cite{srivastava2019data}, authors explore data sharing and privacy for patient IoT devices using block-chain. In \cite{luo2016unlinkable}, the secure storage of NFC generates the key pair (public, private) for a virtual account, and a private key signs all messages during the transaction process, which ensures non-repudiation in their scheme. 

\begin{table}[!h]
\centering
\caption{Security specific categorization of reviewed research papers}
\label{overallsecutirty}
\begin{tabular}{|l| p{4cm}|l|}
\hline
\textbf{Security feature} & \textbf{Papers} & \textbf{\%} 
\\ \hline

Confidentiality & 
 \cite{manvi2009secure, liu2005system, chen2010nfc, al2017online, ruan2014desgn, singh2012comparative, chen2016nfc, purnomo2016mutual, ma2015design, harb2008securesmspay, rui2015design, fan2017u2f, sureshkumar2017lightweight, kang2017privacy, luo2016unlinkable, bojjagani2017secure, khan2016modified} & 80.09 
\\ \hline 
Authentication & 
 \cite{qadeer2009novel, manvi2009secure, liu2005system, zheng2003study, nseir2013secure, chen2010nfc,al2017online, chen2016nfc, purnomo2016mutual, ma2015design, harb2008securesmspay, ruan2014desgn, singh2012comparative, firoz2017defensive, sureshkumar2017lightweight, rui2015design, fan2017u2f, kang2017privacy, luo2016unlinkable, bojjagani2017secure, khan2016modified} & 100 
\\ \hline 
Integrity & 
 \cite{chen2010nfc, al2017online, chen2016nfc,purnomo2016mutual, ma2015design, harb2008securesmspay, rui2015design, fan2017u2f, kang2017privacy, luo2016unlinkable, bojjagani2017secure, khan2016modified} & 57.14 
\\ \hline 
Mutual Authentication & \cite{liu2005system, al2017online, chen2016nfc, purnomo2016mutual, ma2015design, firoz2017defensive, rui2015design,fan2017u2f, kang2017privacy, luo2016unlinkable, bojjagani2017secure, khan2016modified} & 57.14 
\\ \hline 
Customer Anonymity & \cite{liu2005system, luo2016unlinkable, kang2017privacy, sureshkumar2017lightweight, singh2012comparative, chen2010nfc,chen2016nfc} & 33.33 
\\ \hline 
Non-Repudiation & 
 \cite{liu2005system, al2017online, ma2015design, harb2008securesmspay, rui2015design, fan2017u2f, luo2016unlinkable, bojjagani2017secure} & 38.09 
\\ \hline
\end{tabular}
\end{table}

The overall security features provided by each paper in our study are described in Table \ref{overallsecutirty}. It tells us that all the systems we have reviewed ensure authentication, and most of them also provide encryption. The main aspects considered in each payment system are encryption and authentication; without these two aspects, no system can be said as secure enough. Integrity and registration of clients or merchants have also got much importance and value while designing any payment system.

\section{Challenges and Future Work}\label{Sec6}

Due to the increase in technology used worldwide to ease daily life activities, mobile payment systems also emerged rapidly for the same reasons. Tasks that take hours to perform by visiting the banks are now at the fingertips using smartphones and allied payment infrastructure in digital forms. This ease also brought some related issues, the most dangerous of which is the threat of malicious actors hacking the payment system to steal money. The recent hacking of block-chain-based cryptocurrency exchanges, which were previously considered the most secure digital payment system, rings the bells that the hackers circumvent ways to bypass the securities in place. This new battleground between the good and the bad for enhancing and ensuring the security of mobile payment systems against emerging threats is an affluent area to explore in the future.

In any field, there is always the possibility of enhancements and improvement. In the future, we intend to focus on understanding the preferences of consumers and the reasons to utilize or not utilize a specific technology-enabled service as it is vital to design viable services that generate value to consumers and the other stakeholders of an ecosystem. The usage of mobile phones is high, and it is in almost every person's approach. Most of the work or daily transactions or communication is done through a mobile phone; that is why many companies introduced their services for mobile phones. Mobile payment methods are also available nowadays, but it needs more security than other mobile phone services. 

An increase in mobile payment solutions will increase the user base, which is already sufficient compared to other traditional methods. This increase will ultimately result in a load on the network infrastructure, which is the backbone of the success of such solutions. Advancement in next-generation networks and their impact on mobile payment solutions will be another research area to explore. Further to this, research can be done on current bottlenecks resulting in lesser mobile payment solutions and remedial measures using network advancements. 

This research has some practical and theoretical limitations that may provide valuable findings for future research. For example, we do not consider the potential impact of digitization on mobile payment systems, making behaviors more complex than those resulting from modular reorganization alone. Our goal when choosing this project is to record dynamics that cannot be found in developed countries. We hope our findings can be applied to other mobile payment systems in emerging economies. However, future comparative studies using larger samples or more extreme cases will confirm the extent to which our results can be generalized. Since all cases are based on mobile network operators (MNOs), future research on banks or third-party models will help discuss mobile payment systems in the literature.

\section{Conclusion}\label{Sec5}
This paper has discussed various payment schemes and their usage, technology, and provided security. Most payment methods are account-based payment systems, and their main focus is on security, privacy, confidentiality, and authentication. We present an overview and discussed different components of MPS. We present a detailed survey of the existing MPS structure and its limitations; provide detailed history, development, and deployment of MPS. Discussed different aspects of MPS included socioeconomic conditions, Cost Efficiency, diffusion of mobile phones, convenience, new initiatives, heavy restrictions and regulations, limited collaboration, underdeveloped ecosystem, and security problems; the key attributes of MPS, and stakeholder and communication entities roles in MPS form different aspects. We discussed different security mechanisms involved in MPS. Provide analysis of the encryption technologies, authentication methods, and firewall in MPS. All the papers suggest different techniques to provide different security aspects. However, the main point is that keeping in check the CIA triad, each payment should be made with authentication and encryption because the future of MPS depends on its security features.

\bibliographystyle{ieeetr}
\bibliography{ref}
\EOD

\end{document}